\documentstyle[prb,preprint,aps,psfig]{revtex}
\begin{document}
\draft


\title{Chlorine adsorption on the Cu(111) surface}
\author{K. Doll$^+$ and N. M. Harrison$^*$}
\address{CLRC, Daresbury Laboratory, Daresbury, Warrington, WA4 4AD, UK}
\address{$^*$ 
and Department of Chemistry, Imperial College, London, SW7 2AY, UK}
\address{$^+$ Electronic address: k.doll@dl.ac.uk}
\maketitle

\begin{abstract}
We investigate the adsorption of chlorine on the Cu(111) surface with
full potential all-electron density functional calculations. 
Chlorine adsorption at the fcc hollow sites is slightly preferred over that
at the hcp hollow.
The adsorption geometry is in excellent agreement with electron diffraction
and ion scattering data. 
Adsorption energies and surface diffusion barriers are close to those deduced 
from experiment.
\end{abstract}

\pacs{ }

\narrowtext
\section{Introduction}

The determination of surface structure and adsorption energetics
is a key issue in surface science \cite{Zangwill}. Adsorption on 
metallic surfaces is of particular importance due to its 
relevance to many industrial processes. First principles 
simulation, based on density functional theory, is now one of 
the key tools for studying and developing an understanding of 
these systems \cite{SchefflerStampfl}.

The adsorption of halides on metal surfaces and especially Cl on Cu(111) 
has been the subject of extensive experimental studies. The first
study in 1977 reported on results from 
low energy electron diffraction
(LEED), Auger electron spectroscopy (AES),
work function measurements and desorption experiments \cite{GoddardLambert}.
Further work has been performed using surface extended X-ray absorption fine
structure (SEXAFS) and photoelectron diffraction
\cite{Crapper1986,Crapper1987}, normal
incident X-ray standing wavefield absorption (NIXSW) 
\cite{WoodruffPRL1987,Woodruff1988,Luedecke,Kadodwala}, secondary
ion mass spectrometry (SIMS) \cite{Way},
scanning tunneling microscopy (STM) \cite{Motai},
LEED, AES, thermal and electron stimulated desorption \cite{Walteretal}.
However, we are not aware of any first principles simulations of this 
system. This is no doubt in part due to the large computational effort
needed to study the complex surface geometry with sufficient accuracy.
In this article we present the results of an extensive study of the
surface adsorption geometry and energetics within a gradient-corrected
density functional formalism. It turns
out that a very high accuracy is necessary to distinguish the different
adsorption sites and thus careful calculations
capable of resolving energy differences of the order of only a few meV
are required. In view of this we present details of the tests
performed to establish the numerical accuracy of the calculations.

The outline of the article is as follows. In section 
\ref{Basissetsection}, we give
the details of the computational method. In section \ref{Copperbulk},
we present results for bulk Cu which document the accuracy of the
numerical methodology and the functional used. In section
\ref{Coppersurfaces}, the surface geometry and energetics of the clean
Cu surface are presented. The important numerical approximation of
reciprocal lattice sampling is discussed in detail.
In section \ref{Chlorineoncopper} we present a quantitative description
of the adsorption of Cl on the (111) surface of Cu.

\section{Method}
\label{Basissetsection}

The fundamental approximation made in density functional calculations
is the choice of the effective one-particle potential. In this study
we employed three distinct approximations: the local density approximation
(LDA) with Dirac-Slater exchange \cite{LDA} 
and the Perdew-Zunger correlation functional \cite{PerdewZunger};
the gradient corrected exchange and correlation functional of Perdew and Wang
(GGA) \cite{Perdewetal}; and the Hartree-Fock (HF) approximation.

A local basis scheme based on atom-centred Gaussian type orbitals
was used. The calculations were performed with the CRYSTAL98 
software \cite{Manual}.
The Cu basis set was chosen as $[6s5p2d]$ \cite{KCuF3}. 
The values of the inner exponents 
(a set consisting of one $s$ contraction, 3 $sp$ contractions and 
1 $d$ contraction, i.e. $[4s3p1d]$), is given in table 
\ref{CRYSTALbasis}. The outermost 
exponents were optimized in calculations on copper bulk at the 
GGA, LDA, and HF level. It is interesting to note that for Cu, in
contrast to lithium \cite{Dovesi}, it is  possible to optimize 
the total energy with respect to the exponents without numerical
difficulties even in the HF approximation. 
We believe that this is due to the more localized
nature of the orbitals in Cu.

For Cl the $[5s4p]$ 
basis set developed in previous calculations on alkali halides was used
\cite{Prencipe}, with outermost $sp$ exponents of 0.294 and 0.090. An
additional $d$-function with exponent 0.5 was added yielding a
$[5s4p1d]$ set. 

The density functional potential was fitted with an auxiliary basis set,
which consisted, for both Cu and Cl, of 12  $s$ and $p$ functions with
exponents taken to be a geometrical sequence from  0.1 to 2000,
and 5 $d$ functions with exponents in the range 0.8 to 100. 

As in a recent study of lithium
\cite{KlausNicVic}, we employ finite temperature density functional
theory \cite{Mermin} to ease the numerical integration over $\vec k$-space
with the occupation
calculated according to the Fermi function at finite temperature 
 $T$ (Ref. \onlinecite{FuHo})
 and the zero temperature energy is approximated by
\cite{Gillan}
$\frac{1}{2}(E(T)+F(T))$; with $E$ the energy and $F$ the free energy 
$F=E-TS$, where $S$ is the electronic free entropy.
The important approximation associated with the choice of  the 
sampling nets employed in reciprocal space integration will be
discussed in detail below. 

\section{Bulk properties}
\label{Copperbulk}

We calculated the band structure and cohesive properties
of Cu bulk in order to document the performance of our
numerical approximations and to compare different choices for
the effective one-particle potential (LDA, GGA, and HF).

Figure \ref{cuLDAband} displays the LDA band structure for bulk copper which is
in excellent agreement with that reported in the 
literature \cite{LDAbandliterature}. The LDA band structure remains 
virtually unchanged when only LDA exchange is included. Using GGA exchange
and correlation leads to a similar band structure.
The Hartree-Fock band structure (figure \ref{cuHFband})
reveals the expected pathology associated with the non-local
exchange interaction and neglect of correlation in a metal.
The overall bandwidth is too large by a factor of nearly two
and the $d$-bands too low in energy compared to the wide 4$sp$ 
band. 

The structure and cohesive properties are reported in 
table~\ref{Coppergroundstatetable}. The GGA cohesive
energy, lattice constant and bulk modulus are in good 
agreement with those observed. The LDA tends to over-bind
the system resulting in a somewhat too low lattice constant
and thus a high bulk modulus. The HF solution is dramatically
underbound, the lattice constant far too large and thus
the bulk modulus is very low. It is clear that the structure and
energetics of the bulk crystal are best described within the
GGA.

\section{Cu surfaces}
\label{Coppersurfaces}

We studied the clean Cu surface in some detail in order
to establish the accuracy of predicted surface structures
and energies.

We used two methods to compute the surface 
energy \cite{BoettgerTrickey1992,KlausNicVic}.
Firstly, by extrapolating from calculations on slabs
with a different number of layers ($n$ and $m$):

\begin{equation}
\label{nn-1gleichung}
E_{surface}=
\frac{1}{2}(E_{slab}(n)-(E_{slab}(n)-E_{slab}(n-m))\frac{n}{m})
\end{equation}

and secondly by using an independent bulk energy:

\begin{equation}
\label{eebulkgleichung}
E_{surface}=\frac{1}{2}(E_{slab}(n)-E_{bulk} \times n)
\end{equation}

where all the quantities $E_{surface}$, $E_{slab}(n)$, 
and $E_{bulk}$ are expressed as energies per atom. 
In Ref. \onlinecite{KlausNicVic},  we gave 
a detailed comparison of the application of both formulas  to the Li
surface. The electronic structure of Cu is more complex
than Li so as a first step we investigated
the dependence of the surface energies on the electronic
temperature and number of sampling points.

We focus the discussion now on the Cu (111) surface.
In figure \ref{cu3-4layerfigure}, we display the
temperature and sampling point dependence of 
the surface energy obtained from 
equation \ref{nn-1gleichung} using two slabs with 3 and 4 layers.
The density of reciprocal lattice points is determined by a shrinking factor.
We used different shrinking factors of 2, 4, 8, 12, and 16 which
result in 2, 4, 10, 19, and 30 points in the irreducible part of
the Brillouin zone. As expected the surface energy converges with 
respect to $\vec k$-point sampling at higher temperatures but as the
temperature is raised the extrapolation to the athermal limit
becomes less accurate.

For the very high accuracy required in the current study we
selected a shrinking factor of 16 and a temperature of $k_BT = 0.01 E_h$ 
($E_h$=27.2114 eV). This converges the total energy of the bulk
to better than 0.1 $mE_h$, and that of the three layer copper slab 
to better than 0.3 $mE_h$.
In table \ref{surfenergyconvtable}, we show the variation of 
the computed
surface energy with slab thickness obtained both with equation  
\ref{nn-1gleichung} and \ref{eebulkgleichung}. The convergence 
of the data based on equation \ref{eebulkgleichung} demonstrates
the independent convergence of the bulk and slab energies with
respect to $\vec k$-space sampling. This is an important prerequisite
for accurate studies of chemical processes at surfaces. In many surface
studies bulk cells and $\vec k$-space sampling are chosen to eliminate
systematic errors in the definition of the surface energy. We note
that this is not sufficient to guarantee the accuracy of surface
properties.

Table 
\ref{Coppersurfsummtable} contains the computed surface energies  of
copper. As found in earlier studies on the homogeneous electron
gas\cite{Perdewetal} and lithium \cite{KlausNicVic}, the LDA calculation
results in a higher surface energy than the GGA. 
Previous LDA results \cite{Rodach,Skriver} are rather scattered but
in reasonable agreement with those computed here.

In addition, we studied the relaxation of the  Cu (111)
surface. For this purpose, the top layer of various
slabs of different thickness was allowed to relax
with the layer spacing within the slab fixed at
the bulk value of $\frac { 3.63 {\rm \AA}}{\sqrt{3}}$.
The results in table \ref{Copperrelaxgeomtable} demonstrate that 
convergence is achieved for slabs with 4 or more layers.
The computed inwards relaxation of 1.0 \% is in very good agreement
with recent experiments using medium-energy ion scattering (MEIS) \cite{Chae} 
where a contraction of 1.0$\pm$0.4\% was measured
for the first interlayer spacing; it also agrees with 
that deduced from LEED  experiments where inwards relaxations of 
0.3$\pm$1\% \cite{Tear} and 0.7$\pm$0.5\% \cite{Lindgren}
have been reported. A previous LDA calculation reported
a relaxation of \\-1.27\% \cite{Rodach}.
In this calculation the second interlayer spacing, 
which we kept fixed at the bulk value, was also reported to
decrease by 0.64 \% while an experimental study \cite{Chae} found a
decrease of 0.2 \%.
We find that the relaxation of the top layer reduces the 
surface energy very slightly (by 0.003 $\frac{J}{m^2}$).

\section{Chlorine adsorption on the Cu(111) surface}
\label{Chlorineoncopper}

The adsorption of chlorine on the Cu(111) surface was first studied
by Goddard and Lambert \cite{GoddardLambert}. At a coverage of 
one third of a monolayer, a structure displaying long range order in
a $\sqrt{3} \times \sqrt{3}\ {\rm R}30^\circ$ 
pattern was observed and a desorption 
energy of 236 kJ/mol was obtained (see also Ref. \onlinecite{Walteretal}). 
More recently the $\sqrt{3} \times \sqrt{3}\ {\rm R}30^\circ$ pattern 
has been confirmed
in STM images \cite{Motai}. The distance between the Cl layer and the 
surface  Cu layer has also been deduced from 
SEXAFS and photoelectron diffraction experiments
\cite{Crapper1986,Crapper1987}. In this study, a 
distance of 2.39 $\pm$ 0.02 \AA \ 
between nearest Cl-Cu neighbours was obtained
(corresponding to an interlayer spacing of 1.88 $\pm$ 0.03 \AA \mbox{ }); 
the adsorption site was identified as the fcc (face centered cubic)
hollow (see figure \ref{fort34-figure})\cite{DLV}.
A distance of 1.81 $\pm 0.05$ \AA \mbox{ } between the Cl layer and 
a notional Cu-layer corresponding to unrelaxed bulk termination was deduced
from NIXSW data
\cite{WoodruffPRL1987,Woodruff1988,Luedecke}. Comparing the latter two results
(SEXAFS and NIXSW)
gives evidence for a slight inwards relaxation of the surface Cu layer
similar to that observed on the clean surface.
SIMS measurements~\cite{Way} yield an interlayer spacing
 of 1.87 $\pm$ 0.04 \AA \mbox{ }.
Recently, in NIXSW measurements, the fcc site was confirmed as the
adsorption site, but also a small 
occupation of the hcp (hexagonal closed packed) hollow was observed
\cite{Kadodwala}.

We optimized the structure of a slab with a chlorine
coverage of one third of a monolayer in a $\sqrt{3} \times \sqrt{3}\ 
{\rm R}30^\circ$ cell. GGA calculations with the
computational parameters described in section \ref{Coppersurfaces}
were performed
which led to 73 sampling points in the irreducible part of the
reciprocal lattice.
Slabs consisting of 3 and 4 layers of Cu were used in which the Cu
atoms in the top layer and the Cl layer 
were allowed to relax perpendicular to the surface. 
Cl was adsorbed in a number of sites;
the threefold hcp hollow, the threefold fcc hollow, a top position, and
a bridge position (figure \ref{fort34-figure}).

The relaxed geometries and adsorption energies are presented 
in table
\ref{ClonCutable}. Adsorption on the fcc or hcp sites is clearly preferred
over adsorption on bridge or top sites (by  3 and 17 $mE_h$, respectively). 
The Cl-Cu interlayer distance is 1.89 \AA \mbox{ } for
the fcc hollow (1.90 \AA \mbox{ } for the hcp hollow) in excellent
agreement with that deduced from experiment. The next neighbour Cl-Cu 
distance is
2.40 \AA \mbox{ } for the fcc site, 2.41 \AA \ for the
 hcp site, 2.33 \AA \mbox{ } 
for the bridge site and 2.17 \AA \mbox{ } for  the top
site. This is consistent with the idea that the next-neighbour
bond strength is greater, and therefore the bond shorter,
when fewer nearest neighbours
are available \cite{Pauling}.
The inwards relaxation of the copper layer of
$\sim$ 0.04 \AA \mbox{ } (1.9 \%) is insensitive to the Cl adsorption
site and similar to that found for the clean surface
(1.7 \%, table \ref{Copperrelaxgeomtable}).

The energy difference between the fcc and hcp sites is
0.3 $mE_h$ and increases to 0.4 $mE_h$ when the Cu surface is relaxed.
To investigate the dependence of this rather delicate result
 on the number of layers
of the slab, we performed similar calculations for fcc and hcp
site on a four layer slab.
The fcc site remains the preferred one by 
0.2 $mE_h$. The relaxed surface geometry (i.e. the interlayer
distance Cl-Cu) is identical to that
computed for the three layer slab.
The relaxation of the top Cu layer reduces slightly to 0.03 \AA (1.4 \%) 
which is a similar trend as for the clean surface (table 
\ref{Copperrelaxgeomtable}). This value is within the
errorbars of the experimentally expected value.
The optimization indicates that the
fcc site is becoming more stable
relative to the hcp site as the interlayer spacing between Cl and the
second Cu layer is reduced.
This is demonstrated in Figure \ref{hcpminusfccfigure} where,
at a fixed interlayer spacing 
of 1.886 \AA \mbox{ } between Cl and the top Cu layer, the
top Cu layer is allowed to relax inwards. As the relaxation increases, the fcc
site becomes more stable with respect to the hcp site. 
The small energy difference between
the fcc and hcp adsorption sites is in
agreement with observations of predominantly fcc-site adsorption
accompanied by a small occupancy of the hcp site \cite{Kadodwala}.

The calculated adsorption energy 
of 0.135 $E_h$ (per Cl atom) is in reasonable agreement with
that deduced from desorption measurements 
($\sim$ 250 kJ/mol corresponding to 0.095 $E_h$)
\cite{GoddardLambert,Walteretal}. The experimental desorption energy 
is obtained from an Arrhenius model and therefore depends on
a pre-exponential factor which was chosen to be 10$^{13}\frac{1}{s}$
(Ref. \onlinecite{GoddardLambert,Walteretal}),
but questioned in Ref. \onlinecite{GoddardLambert}.

The activation energy for surface diffusion was measured 
experimentally by electron stimulated diffusion\cite{Walteretal}, i.e. the
activation energy for Cl atoms to diffuse into
an area where Cl depletion  is induced by an electron beam.
A value of 19 kJ/mol or 7 $mE_h$ was measured.
A crude estimate from our calculations is possible if we assume that
the energy difference of 3 $mE_h$ between the hollow site and the
bridge site is of the order of the activation energy for diffusion.

\section{Conclusion}
We have demonstrated that a quantitative description of the
adsorption of Cl on the Cu(111) surface can be achieved with
full potential, all-electron, GGA calculations. The surface
structure and adsorption energies are in 
excellent agreement with experiment. 
The Cl-Cu bond length is found to be 1.89  \AA \mbox{ } and
the top Cu layer relaxes inwards by 1.4 \%. 
The fcc hollow is found to be the preferred adsorption site, with
the hcp site being $\sim$ 0.2 $mE_h$ (5 meV) higher in energy.
The energy difference between the two sites is sufficiently small
that it may alter slightly if a larger slab geometry was considered.
Similarly, by analogy to the results for the clean Cu slab we  might
expect the inwards relaxation of 1.4 \%  to become slightly lower with
larger slab geometries. 
The adsorption energy and surface diffusion energy
is in reasonable agreement with those which can be estimated from
experiment.
Bulk and surface properties
of Cu metal are found to be in very good agreement with experiment
and previous calculations, when the gradient corrected GGA functional is used.

\section{Acknowlegments}
The authors would like to acknowledge support from EPSRC grant
GR/K90661.

\onecolumn

\newpage
\begin{table}
\begin{center}
\caption{\label{CRYSTALbasis} The $[6s5p2d]$ copper basis set with
the two  outermost $sp$ exponents and the outermost $d$ exponent
 optimized within the GGA.
Those based on LDA and HF are shown in brackets.}
\vspace{5mm}
\begin{tabular}{ccccc} 
& exponent & $s$ contraction & $p$ contraction & $d$ contraction \\
$s$ &   398000.0  &  0.000227\\
       &    56670.0  &  0.001929\\
       &    12010.0  &  0.01114\\
       &     3139.0  &  0.05013\\
       &      947.2  &  0.17031 \\
       &      327.68 &  0.3693 \\
       &      128.39 &  0.4030 \\
       &       53.63 &  0.1437 \\
\\
$sp$ &  1022.0  &  -0.00487 &  0.00850\\
  &   238.9  &  -0.0674  &  0.06063\\
  &    80.00 &  -0.1242  &  0.2118 \\
  &    31.86 &   0.2466  &  0.3907 \\
  &    13.33 &   0.672   &  0.3964 \\
  &     4.442 &  0.289   &  0.261  \\
\\
$sp$   &    54.7   &  0.0119  & -0.0288 \\
  &    23.26  & -0.146   & -0.0741 \\
  &     9.92  & -0.750   &  0.182  \\
  &     4.013 &  1.031   &  1.280  \\ \\
$sp$ &      1.582 &  1.0  &     1.0\\ \\
$sp$ &      0.596 (LDA: 0.610; HF: 0.555) &  1.0  &     1.0\\ \\
$sp$ &      0.150 (LDA: 0.150; HF: 0.170) &  1.0  &     1.0\\ \\
$d$  
    &  48.54 & & & 0.031 \\
    &  13.55  & & & 0.162 \\
    &  4.52  & & &  0.378 \\
    &  1.47   & & & 0.459 \\ \\
$d$  &    0.392 (LDA: 0.392; HF: 0.423)  & & & 1.0\\
\end{tabular}
\end{center}
\end{table}

\newpage
\begin{table}
\begin{center}
\caption{\label{Coppergroundstatetable}
The ground state properties of bulk copper. Energies
are in Hartree units, lattice constants in \AA, bulk moduli in GPa.}
\vspace{5mm}
\begin{tabular}{ccccccc}
 & &  &  & \\
 & $a$ & $E_{coh}$  & $B$ \\
LDA  
 & 3.53 & 0.182 & 195 \\
GGA 
 & 3.63 & 0.143 & 155 \\
HF & 3.95 & 0.018 & 69 \\
\\
Lit. (KKR)  \cite{JanakMoruzziWilliams}
& 3.59  & 0.155 & 158 \\
Lit. (LDA) \cite{Rodach} & 3.62 & 0.133 & 147 \\
exp. & 3.604 \cite{Bross} & 0.129 \cite{Gschneider}  & 142\cite{Overton} \\
\end{tabular}
\end{center}
\end{table}

\newpage
\begin{table}
\begin{center}
\caption{\label{surfenergyconvtable} The convergence of the surface energy
of the Cu(111) surface as a function of the number of layers, computed 
using
a shrinking factor of 16 and a smearing temperature of 0.01 $E_h$. For
the Cu(111) surface, 1 $\frac{E_h}{atom}$ corresponds to 76.4 
$\frac{J}{m^2}$ for a lattice constant of 3.63 \AA.}

\begin{tabular}{ccccc}

number & $E_{slab}(n)$ & $E_{slab}(n)-E_{slab}(n-1)$ 
& E$_{surface}$ using & E$_{surface}$ using\\
 of layers & &  & $E_{slab}(n)-E_{slab}(n-1)$ & E$_{bulk}=-1640.698861 $\\
  & $E_h$ & $\frac{E_h}{atom}$ & $\frac{E_h}{atom}$ & $\frac{E_h}{atom}$ \\
   1  & -1640.653606  &       -         &      -        & 0.02263 \\
   2  & -3281.350305  & -1640.69670     & 0.02155       & 0.02371 \\
   3  & -4922.049528  & -1640.69922     & 0.02407       & 0.02353 \\
   4  & -6562.748468  & -1640.69894     & 0.02365       & 0.02349 \\
   5  & -8203.447228  & -1640.69876     & 0.02329       & 0.02354 \\
   6  & -9844.146061  & -1640.69883     & 0.02347       & 0.02355 \\
   7 & -11484.844913  & -1640.69885     & 0.02352       & 0.02356 \\
   8 & -13125.543718  & -1640.69881     & 0.02336       & 0.02359 \\
   9 & -14766.242595  & -1640.69888     & 0.02365       & 0.02358 \\
  10 & -16406.941437  & -1640.69884     & 0.02349       & 0.02359 \\
  11 & -18047.640289  & -1640.69885     & 0.02354       & 0.02359 \\
\end{tabular}
\end{center}
\end{table}

\newpage
\begin{table}
\begin{center}
\caption{\label{Coppersurfsummtable}The surface energy ($\frac{J}{m^2}$) of
the low index copper surfaces.}
\begin{tabular}{ccccccc}
surface & LDA & GGA  & Ref. \onlinecite{Skriver}
(LDA) ; Ref. \onlinecite{Rodach} (LDA) \\
(100) & 2.52 & 2.01 & 2.09 ; 1.712 \\
(110) & 2.67 & 2.15 & 2.31 ; 1.846 \\
(111) & 2.26 & 1.80 & 1.96 ; 1.585 \\
\end{tabular}
\end{center}
\end{table}

\newpage
\begin{table}
\begin{center}
\caption{\label{Copperrelaxgeomtable}Relaxation of Cu (111) surface,
as a function of the number of layers, computed within the GGA. 
The top layer is allowed to relax,
the other layers are kept at a fixed distance of $\frac{3.63 {\rm \AA}}{\sqrt{3}}$
corresponding to the bulk lattice constant}
\begin{tabular}{cc}
number of layers & relaxation in {\rm \AA} \mbox{ } (in \%) \\
3 & -0.035 (-1.7 \%) \\
4 & -0.025 (-1.2 \%) \\
5 & -0.023 (-1.1 \%) \\
6 & -0.022 (-1.0 \%) \\
\\
Literature: & \\
exp.\cite{Chae} & -1.0 $\pm$ 0.4 \% \\
exp.\cite{Tear} & -0.3 $\pm$ 1 \% \\
exp.\cite{Lindgren} & -0.7 $\pm$ 0.5 \% \\
LDA \cite{Rodach} & -1.27 \% \\

\end{tabular}
\end{center}
\end{table}

\newpage
\begin{table}
\begin{center}
\caption{\label{ClonCutable} Adsorption of Cl on the Cu(111) surface. 
$\delta_{1-2}$
is the change in interlayer spacing between first and second copper layers
(in \AA) relative to the bulk value, 
$d_{\rm {Cl-Cu \mbox{ }top\mbox{ } layer}}$ is the  distance between the Cl and
top Cu layer (in \AA). The adsorption energy is the difference 
$E_{\rm {Cl \mbox{ }at \mbox{ } Cu(111)}}-{E_{\rm Cu(111)}-E_{\rm Cl}}$. }
\begin{tabular}{cccc}
fcc site\\
n & $\delta_{1-2}$ & $d_{\rm {Cl-Cu \mbox{ }top\mbox{ } layer}}$ 
& $E_{adsorption}$ in $E_h$,
per Cl atom \\
3 & 0 (no relaxation allowed) & 1.89 & 0.13542 \\
3 & -0.043 & 1.89  & 0.13566 \\
4 & -0.032 & 1.89 & 0.13583  \\
hcp site\\
3   & 0 (no relaxation allowed) & 1.90 & 0.13510 \\
3   & -0.038 & 1.90 & 0.13527 \\ 
4   & -0.031 & 1.90 & 0.13563 \\
bridge site\\
 3 & -0.040 & 1.94 & 0.13265 \\
top site\\
 3 & -0.041 & 2.17 & 0.11888 \\
\end{tabular}
\end{center}
\end{table}

\newpage
\begin{figure}
\caption{LDA band structure at the equilibrium lattice constant.}
\label{cuLDAband} 
\end{figure}

\begin{figure}
\caption{HF band structure at the equilibrium lattice constant.}
\label{cuHFband} 
\end{figure}

\begin{figure}
\caption{(111) Copper surface energy with four different reciprocal lattice
samplings extracted from two slabs with 3 and
4 layers using $\frac{1}{2}(E(T)+F(T))$ and equation \ref{nn-1gleichung}}.
\label{cu3-4layerfigure}
\end{figure}

\begin{figure}
\caption{The structures considered for Cl, 
adsorbed on the Cu(111) surface at one
third coverage, in a $\sqrt{3} \times \sqrt{3}\ {\rm R}30^\circ$ unit cell.
When Cl is adsorbed in an fcc hollow, it sits above a Cu atom in the third
layer (upper left) while in the hcp hollow it is above an atom in the 
second layer (upper right).
The top position is Cl adsorbed vertically above a surface atom
(lower left).
In the bridge position it is vertically above the middle of two
surface atoms (lower right). }
\label{fort34-figure}
\end{figure}

\begin{figure}
\caption{Total energy and energy difference of fcc and hcp adsorption site
as a function of the inwards relaxation of the top Cu layer.
The data in this figure is from 
a four-layer slab at a fixed interlayer spacing between
Cl and top Cu layer  of 1.886 \AA \mbox{ }
for both fcc and hcp adsorption site.}
\label{hcpminusfccfigure}
\end{figure}

\newpage
\centerline{\psfig{figure=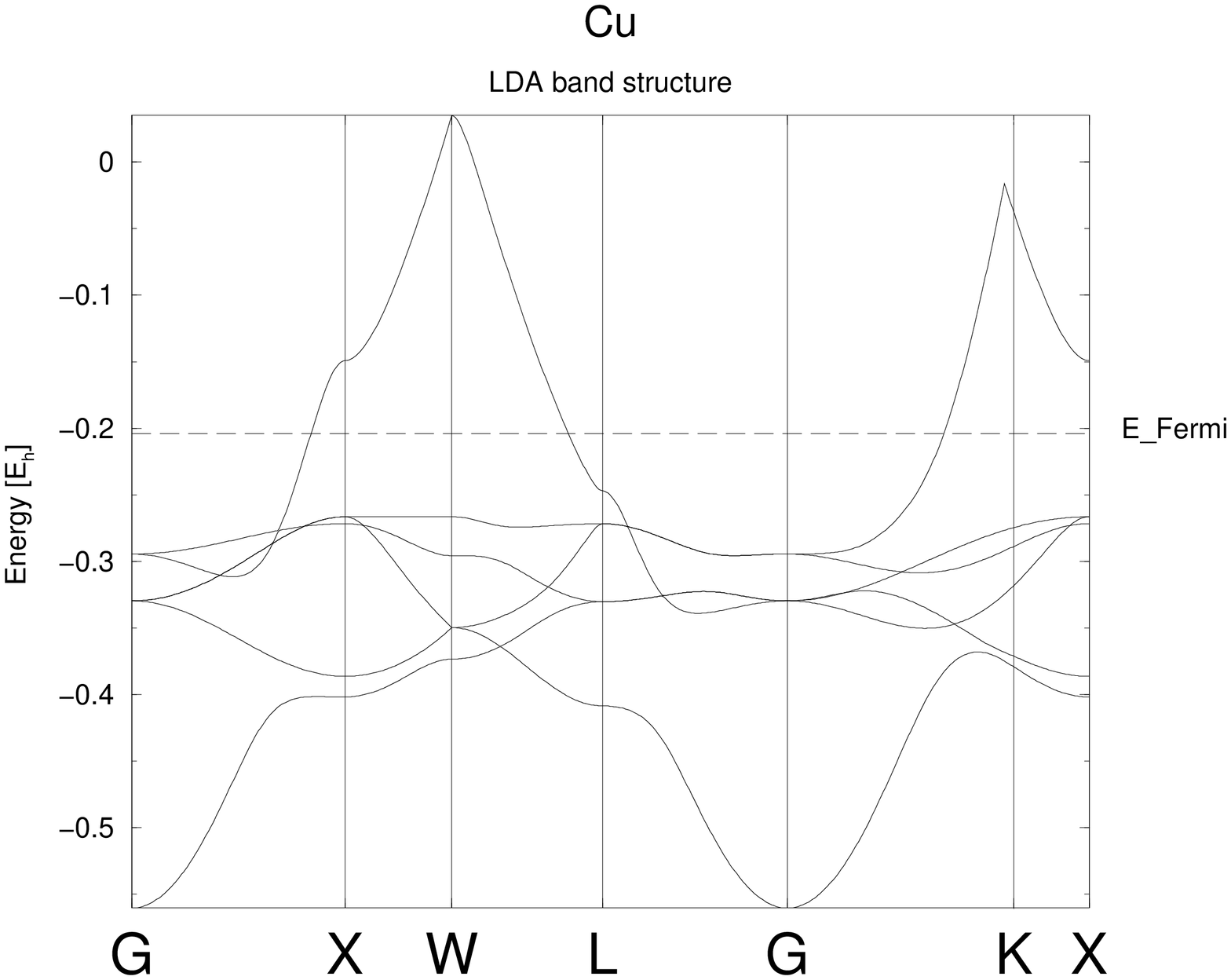,width=15cm,angle=270}}
\vfill

Fig. 1

\newpage
\centerline{\psfig{figure=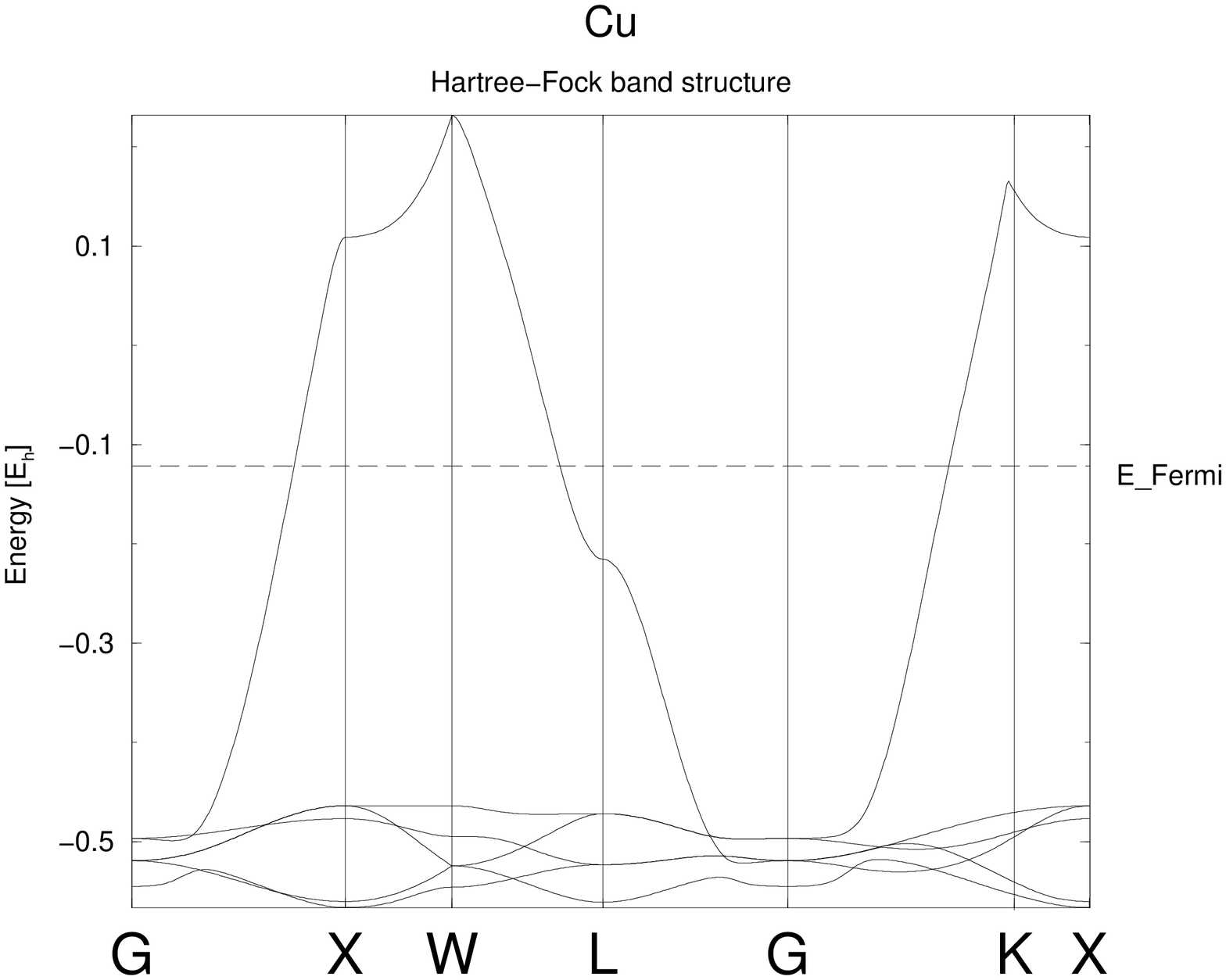,width=15cm,angle=270}}
\vfill

Fig. 2

\newpage
\centerline{\psfig{figure=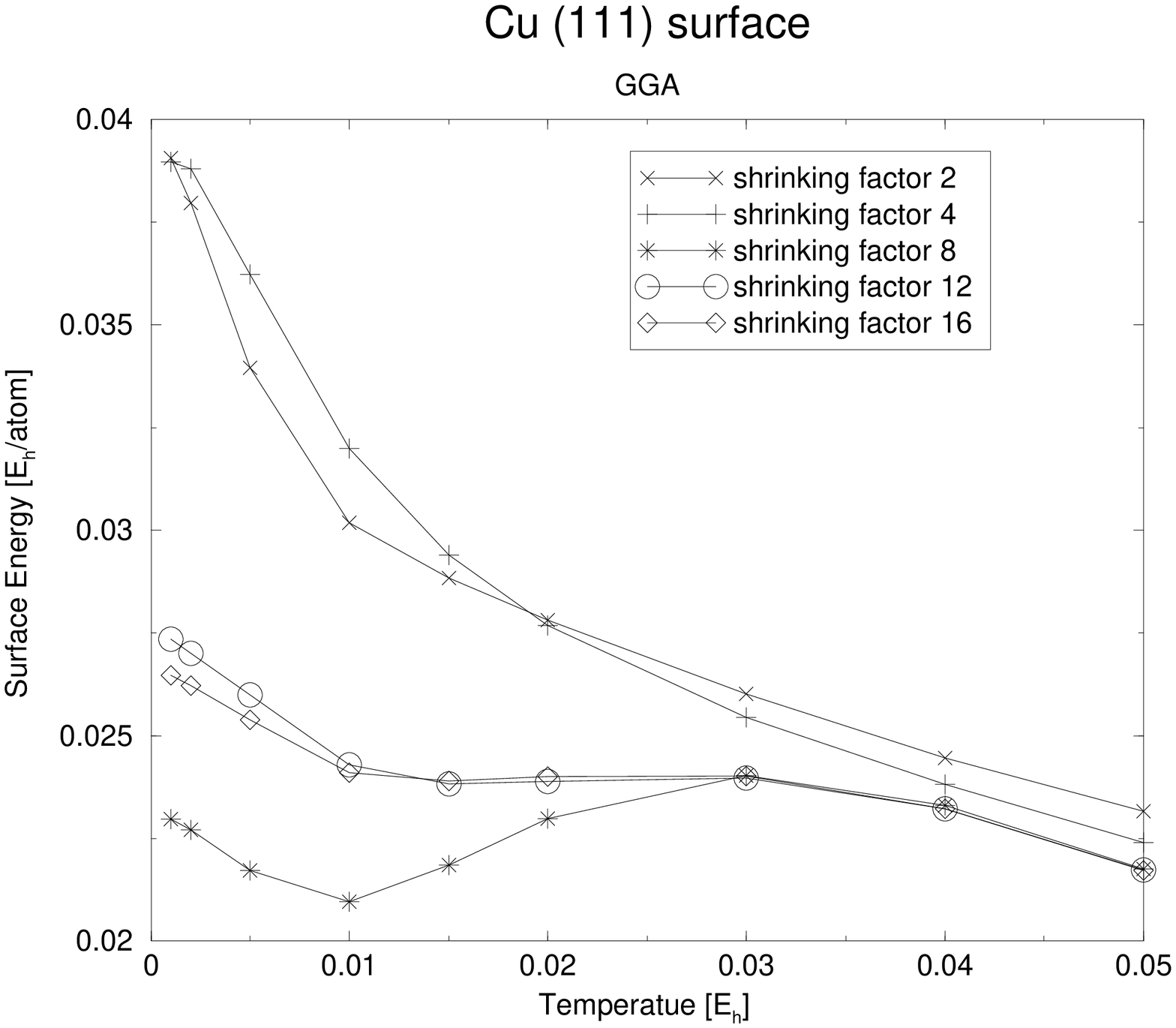,width=15cm,angle=270}}
\vfill

Fig. 3

\newpage
\centerline{\psfig{figure=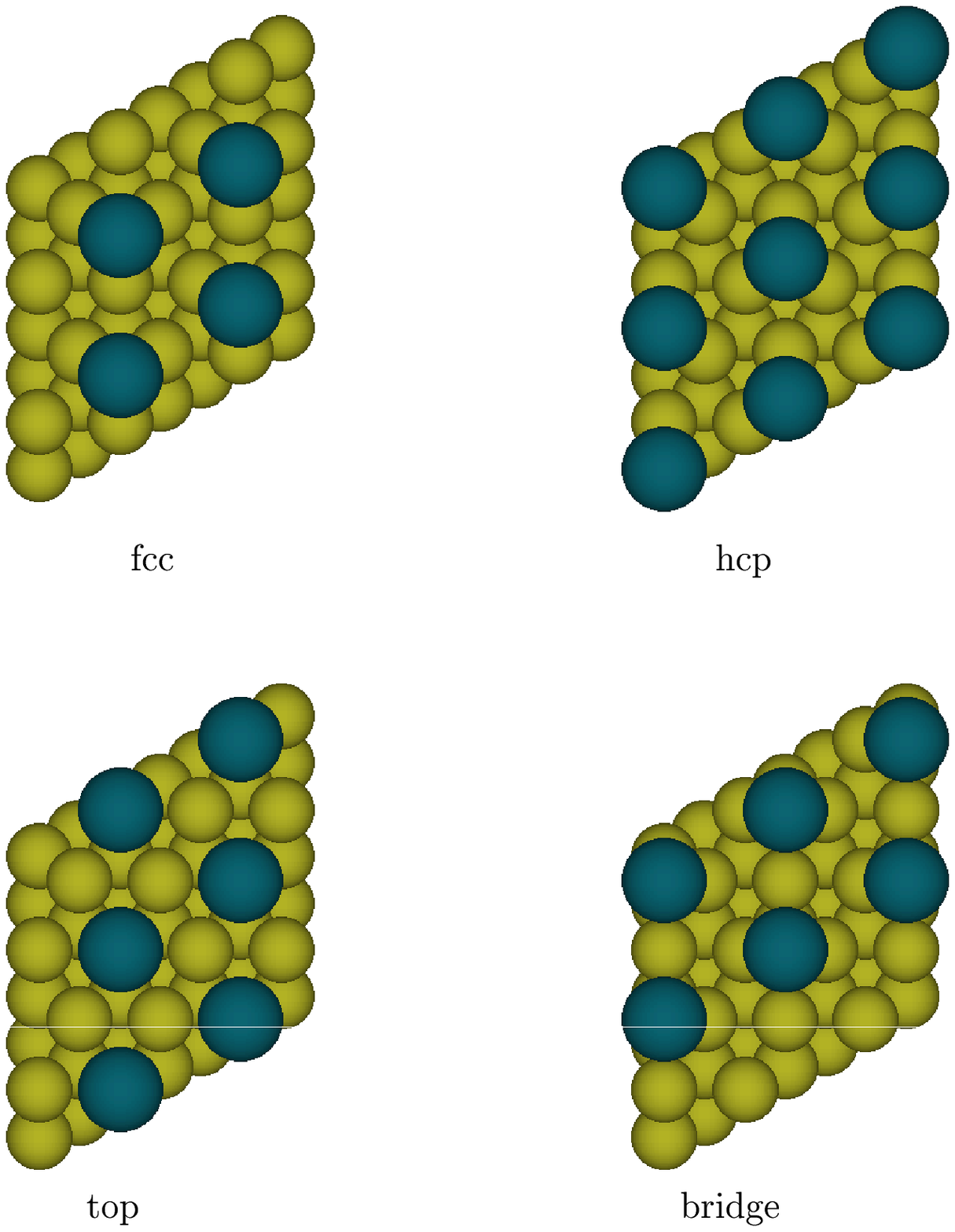,width=15cm,angle=0}}
\vfill

Fig. 4

\newpage
\centerline{\psfig{figure=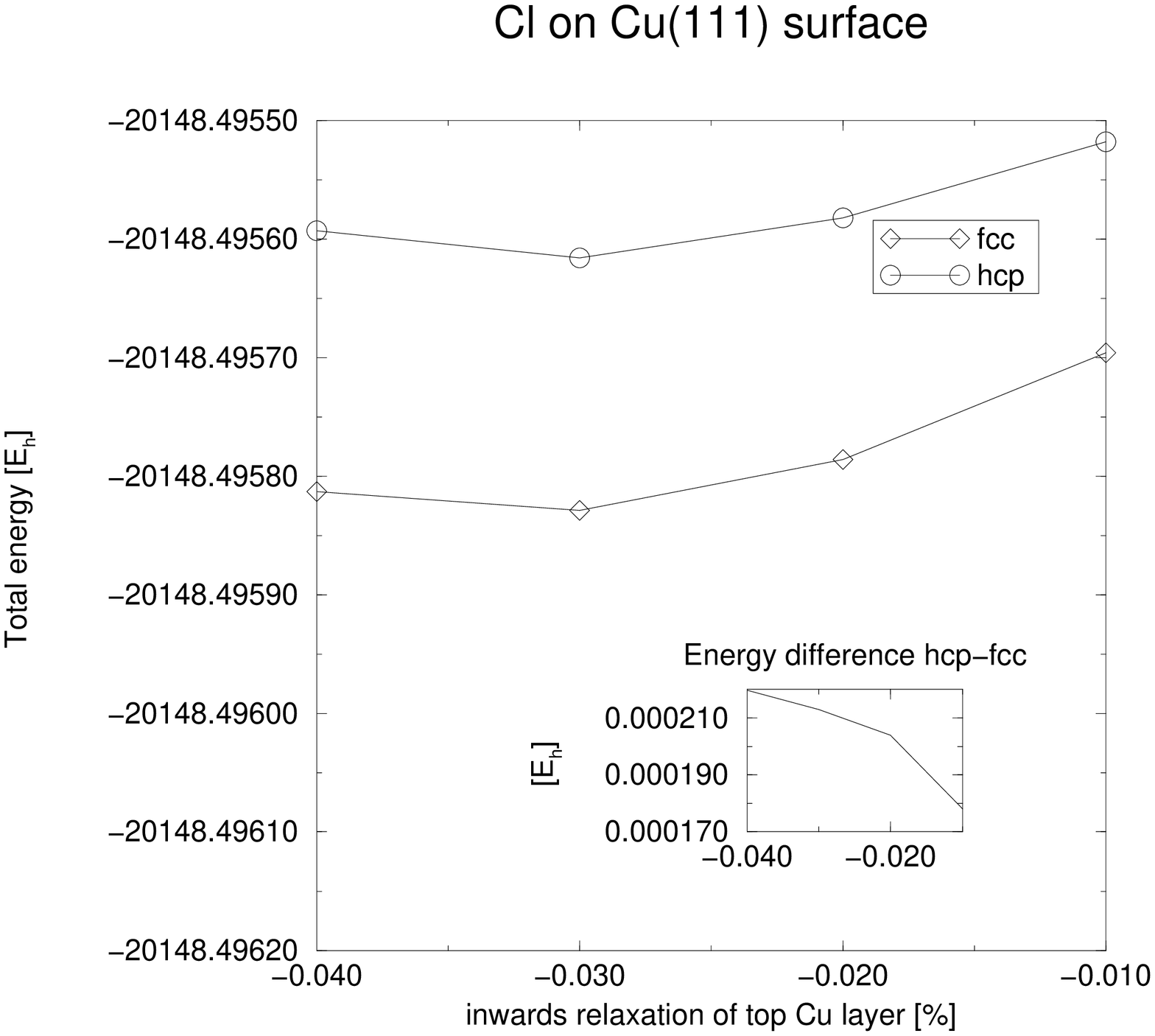,width=15cm,angle=270}}
\vfill

Fig. 5

\end{document}